\documentclass[prb,twocolumn,superscriptaddress,showpacs,amsmath]{revtex4}
\usepackage{graphicx}
\usepackage{dcolumn}
\usepackage{bm}

\def\imth{{\rm i}}

\begin{document}

\title{Mutual Chern-Simons theory for $Z_2$ topological order}

\author{Su-Peng Kou}
\affiliation{\ Department of Physics, Beijing Normal University,
Beijing, 100875 P. R. China }

\author{Michael Levin}

\affiliation{\ Department of Physics, Harvard University, Cambridge,
Massachusetts }

\author{Xiao-Gang Wen}

\homepage{http://dao.mit.edu/~wen} \affiliation{Department of
Physics, Massachusetts Institute of Technology, Cambridge,
Massachusetts 02139}

\begin{abstract}
We study several different $Z_2$ topological ordered states in
frustrated spin systems. The effective theories for those different
$Z_2$ topological orders all have the same form -- a $Z_2$ gauge
theory which can also be written as a mutual $U(1)\times U(1)$
Chern-Simons theory.  However, we find that the different $Z_2$
topological orders are reflected in different projective
realizations of lattice symmetry in the same effective mutual
Chern-Simons theory.  This result is obtained by comparing the
ground-state degeneracy, the ground-state quantum numbers, the
gapless edge state, and the projective symmetry group of
quasi-particles calculated from the slave-particle theory and from
the effective mutual Chern-Simons theories.
Our study reveals intricate relations between topological order and symmetry.

Keywords: topological order, mutual Chern-Simons theory, spin liquid

\end{abstract}

\pacs{75.10.Jm, 11.15.-q, 11.15.Ha}

\maketitle

\section{Introduction}

After the discovery of fractional quantum Hall effect,\cite{TSG8259}, we
realized that new kind of orders beyond Landau's symmetry breaking paradigm is
possible. This new kind order is called topological order \cite{Wrig,Wtoprev}
for gapped states and quantum order\cite{Wqoslpub} for general states. The new
orders reflect patterns of long range entanglements in the ground state.

Gapped $Z_2$ spin liquids have the simplest kind of topological order -- $Z_2$
topological order.\cite{RS9173,Wsrvb} Those topological ordered states may
appear in frustrated spin systems or dimmer
models.\cite{RS9173,Wsrvb,K032,MS0181,BFG0212,MSP0202,Wqoexct} Physically, the
topological orders can be (partially) characterized by robust ground-state
degeneracy.\cite{Wtop,Wsrvb} The low energy effective theory for those $Z_2$
topologically ordered states is a $Z_2$ gauge theory. 

Topological order is a property of a many-body ground state that is robust
against any perturbations, even those perturbations that break all the
symmetries.  In this paper, we like to study the interplay between topological
order and symmetry.  We like to find out how to characterize topological
ordered states that also have certain symmetries.

Recently, it was found that for spin liquids with all the lattice symmetries
(such as lattice translation and rotation symmetry), there can be hundreds
different $Z_2$ topological orders.\cite{Wqoslpub,Wqoexct} We will call those
topological orders symmetric topological orders. It is shown that the
different symmetric $Z_2$ topological orders can be characterized by different
project symmetry groups (PSG).
So those symmetric topological orders are good examples to study
the relation between topological order and symmetry.

Here, we would like to study the low energy effective theories for those
different $Z_2$ topological orders and ask how different symmetric $Z_2$
topological orders are reflected in low energy effective theories. We find
that all different $Z_2$ topological orders can be described by the same
effective mutual $U(1)\times U(1)$ Chern-Simons (CS) theories.\cite{HOS0427}
The lattice symmetry is realized projectively in the effective mutual CS
theories.  It turns out that different symmetric $Z_2$ topological orders have
different projective realizations of the lattice symmetries. To confirm our
results, the projective construction (the slave-particle theory)
\cite{BZA8773,Wen04} is used to calculate the ground-state degeneracies, the
ground-state quantum numbers, and the PSGs of quasi-particles. Those results
agree with those obtained from the effective mutual CS theories. Furthermore,
we also used the effective mutual CS theories to study gapless edge states for
some $Z_2$ topologically ordered states.

\section{Projective construction of many-spin wave functions}

The key to understand topological orders is to construct states that can
have long range quantum entanglements. The projective construction
introduced in the study of high $T_c$ superconductors is a powerful way to
construct such states.\cite{BZA8773,AZH8845,DFM8826,Wen04} In this section,
we will briefly review the projective construction of $Z_2$ topologically
ordered states.

A spin-1/2 model can be viewed as a hard-core-boson model, if we identify
$|\downarrow\>$ state as a zero-boson state $|0\>$ and $|\uparrow \>$ state
as a one-boson state $|1\>$. In the follow we will use the boson-picture to
describe our model.

We first introduce a ``mean-field'' fermion Hamiltonian:\cite{Wqoslpub}
\begin{equation}
H_{\text{mean}}=\sum_{\langle ij\>}\left( \psi _{I,i}^{\dag
}u_{ij}^{IJ}\psi _{J,j}+\psi _{I,i}^{\dag }\eta _{ij}^{IJ}\psi
_{J,j}^{\dag }+h.c.\right) \label{Hmean}
\end{equation}
where $I,J=1,2$. We will use $u_{ij}$ and $\eta _{ij}$ to denote the
$2\times 2$ complex matrices whose elements are $u_{ij}^{IJ}$ and
$\eta_{ij}^{IJ}$.
Let $|\Psi _{\text{mean}}^{(u_{ij},\eta _{ij})}\>$ be the ground state of
the above free fermion Hamiltonian (\ie the lowest energy state obtained by
filling all the negative energy levels). Then a many-boson wave function can
be obtained through
\begin{equation}
\Phi _{\text{spin}}^{(u_{ij},\eta _{ij})}(i_1,i_2...)=\langle
0|\prod_{n=1}^{N_{\text{site}}/2}b(i_n)|\Psi
_{\text{mean}}^{(u_{ij},\eta _{ij})}\>  \label{Phichieta}
\end{equation}
where $N_{\text{site}}$ is the number of lattice sites,
\begin{equation}
b(i)=\psi _{1,i}\psi _{2,i}  \label{bpsi}
\end{equation}
and $i_1$, $i_2$, $\cdots $, label the location of bosons (up-spins). Here,
we have assumed that there are $N_{\text{site}}/2$ up-spins and $N_{\text{
site}}/2$ down-spins.

We may view $(u_{ij},\eta _{ij})$ as variational parameters and the physical
spin 
wave function $\Phi _{\text{spin}}^{(u_{ij},\eta
_{ij})}(i_1,i_2...)$ as a trial wave function. The trial ground
state of a spin Hamiltonian can be obtained by minimizing the
average energy $\langle H\>$.

First let us consider the following spin Hamiltonian
\begin{equation}
H_{\text{exact}}=g\sum_i\hat{F}_i,\ \ \hat{F}_i=\sigma _i^y\sigma _{i+\hat{x}
}^x\sigma _{i+\hat{x}+\hat{y}}^y\sigma _{i+\hat{y}}^x  \label{Hexct}
\end{equation}
where $\sigma ^{x,y,z}$ are the Pauli matrices and $i=(i_x,i_y)$ labels the
site of a square lattice. We find that if we choose the variational
parameters to be
\begin{align}
-\eta _{i,i+\hat{x}}& =u_{i,i+\hat{x}}=1+\tau ^z  \nonumber  \label{Z2Eans}
\\
-\eta _{i,i+\hat{y}}& =u_{i,i+\hat{y}}=1-\tau ^z,
\end{align}
then the spin wave function \Eq{Phichieta} minimize the average energy. In
fact the wave function is the exact ground state of Hamiltonian $H_{\text{
exact}}$. \cite{Wqoexct} It was found that all the excitations above the
ground state are gapped and the ground state contains a non-trivial
topological order described by a $Z_2$ effective gauge theory. We will call
such a state Z2E state.

Ref. \onlinecite{Wsrvb} introduced another many-spin state on square lattice which is
described by
\begin{eqnarray}
u_{i,i+\hat{x}} &=&u_{i,i+\hat{y}}=-\chi \tau ^3,  \nonumber  \label{Z2Aans}
\\
u_{i,i+\hat{x}+\hat{y}} &=&\eta \tau ^1+\lambda \tau ^2,  \nonumber \\
u_{i,i-\hat{x}+\hat{y}} &=&\eta \tau ^1-\lambda \tau ^2,  \nonumber \\
u_{ii} &=&\upsilon \tau ^1.
\end{eqnarray}
and $\eta _{ij}=0$. However, it is not clear what kind of spin Hamiltonian
gives rise to the spin state described by the above variational parameters.
Despite of this, some physical properties of the spin state were obtained
under the assumptions that the state is stable for a certain local spin
Hamiltonian.\cite{Wsrvb} Again, all excitations above the spin state have
finite energy gaps. The spin state is a spin liquid with no spin order. But
it contains a non-trivial topological order described by an effective $Z_2$
gauge theory. So we will call such a spin state Z2A state.

Naively, one may expect the Z2A and the Z2E states to be the same state
since both have $Z_2$ gauge theory as their low energy effective theory. In
the following, we will show that they are different quantum states with
different topological orders.

\section{Ground state degeneracy}

\begin{figure}[t]
\centerline{\includegraphics[scale=0.9]{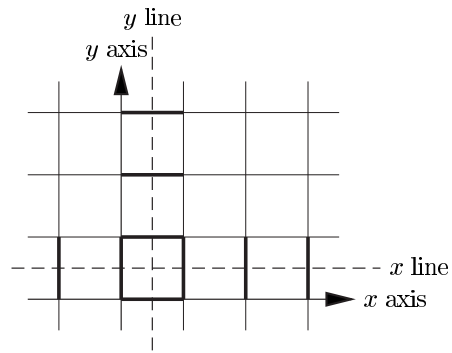} }
\caption{ The links crossing the $x$ line and the $y$ line get an additional
minus sign. }
\label{xyline}
\end{figure}

One way to study a topological order is to study its ground state degeneracy
on a torus. Naively, we expect the Z2A and the Z2E state to have 4
degenerate ground states, as implied by the effective $Z_2$ gauge theory. The
argument goes as the following.

First, we note that the physical boson wave function
$\Phi ^{(u_{ij},\eta_{ij})}(\{i_n\})$
is invariant under the following $SU(2)$ gauge
transformations\cite{Wen04}
\begin{equation}
(\psi _i,u_{ij},\eta _{ij})\to (G_i\psi _i,G_iu_{ij}G_j^{\dag },G_i\eta
_{ij}G_j^T)
\end{equation}
where $G_i\in SU(2)$. So the average energy $E(u_{ij},\eta
_{ij})=\langle \Phi ^{(u_{ij},\eta _{ij})}|H|\Phi ^{(u_{ij},\eta
_{ij})}\>$ satisfies
\[
E(u_{ij},\eta _{ij})=E(G_iu_{ij}G_j^{\dag },G_i\eta _{ij}G_j^T).
\]

Next we assume $(\bar{u}_{ij},\bar{\eta}_{ij})$ give rise to a
(variational) ground state of a Hamiltonian. We would like to show
that the following four ansatz
\begin{align}
u_{ij}^{(m,n)}& =(-)^{ms_x(ij)}(-)^{ns_y(ij)}\bar{u}_{ij}  \nonumber
\label{sl5.1} \\
\eta _{ij}^{(m,n)}& =(-)^{ms_x(ij)}(-)^{ns_y(ij)}\bar{\eta}_{ij}
\end{align}
produce four degenerate ground states. Here $m,n=0,1$. $s_x(ij)$ and $
s_y(ij) $ have values $0$ or $1$. $s_x(ij)=1$ if the link $ij$ crosses the $
x $ line (see Fig.~\ref{xyline}) and $s_x(ij)=0$ otherwise. Similarly, $
s_y(ij)=1$ if the link $ij$ crosses the $y$ line and $s_y(ij)=0$ otherwise.
Physically, the degenerate states arise from adding $\pi $ flux through the
two holes of the torus. The values of $m,n=0,1$ reflect the presence or the
absence of the $\pi $ flux in the two holes.

We note that $(u_{ij}^{(0,0)},\eta _{ij}^{(0,0)})$ represents the
ground state. We also note that $(u_{ij}^{(m,n)},\eta
_{ij}^{(m,n)})$ with different $m$ and $n$ are \emph{locally} gauge
equivalent. This is because, on an infinite system, the change, say,
$u_{ij}\to (-)^{ms_x(ij)}(-)^{ms_y(ij)}u_{ij}$ can be generated by
an $SU(2)$ gauge transformation $u_{ij}\to W_iu_{ij}W_j^{\dag }$,
where $W_i=(-)^{m\Th(i_x) }(-)^{n\Th(i_y)}$, and $\Th(n)=1$ if $n>0$
and $\Th(n)=0$ if $n\leq 0$. As a result,
$E(\bar{u}_{ij},\bar{\eta}_{ij})=E(u_{ij}^{(m,n)},\eta
_{ij}^{(m,n)}) $. On the other hand, on a torus,
$(u_{ij}^{(m,n)},\eta _{ij}^{(m,n)})$ with different $m$ and $n$ are
not gauge equivalent in the global sense. There is no $SU(2)$ gauge
transformation defined on the torus that connects those ansatz. So
the four ansatz give rise to four different degenerate states. This
is how we obtain the four-fold ground state degeneracy for the $Z_2$
states.

However, the above argument is valid only for even by even lattice. For odd
by odd lattice, the argument breaks down. To understand the failure of the
above argument, let us construct the mean-field ground state more carefully.

Let us start with a simple case of the Z2A state. For the ansatz
\Eq{Z2Aans} , the ``mean-field'' Hamiltonian in momentum space
becomes
\begin{widetext}
\begin{eqnarray}
H_\text{mean}(\mathbf{k}) &=&\sum_{\mathbf{k}}(\psi
_{1\mathbf{k}}^{\dagger },\psi _{2\mathbf{k}}^{\dagger })M\left(
\begin{array}{l}
\psi _{1\mathbf{k}} \\
\psi _{2\mathbf{k}}
\end{array}
\right)
=\sum_k\varepsilon (\mathbf{k})\alpha _{\mathbf{k}}^{\dag }\alpha _{\mathbf{k}
}-\sum_k\varepsilon (\mathbf{k})\beta _{\mathbf{k}}^{\dag }\beta
_{\mathbf{k}}
\end{eqnarray}
where
\begin{eqnarray*}
M &=&2\chi (\cos k_x+\cos k_y)\tau ^3+(2\eta \cos (k_x+k_y)+2\eta \cos
(k_x-k_y)+\upsilon )\tau ^1
+(2\lambda \cos (k_x+k_y)-2\lambda \cos (k_x-k_y))\tau ^2,
\end{eqnarray*}
and
\[
\varepsilon (\vec{k})=\sqrt{
4\chi ^2(\cos k_x+\cos k_y)^2+(2\eta \cos (k_x+k_y)+2\eta \cos
(k_x-k_y)+\upsilon )^2
+(2\lambda \cos (k_x+k_y)-2\lambda \cos (k_x-k_y))^2
}.
\]
\end{widetext}
Here $\alpha _{\mathbf{k}}$ and $\beta _{\mathbf{k}}$ are diagonalized
quasiparticles operators
\begin{eqnarray*}
\alpha _{\mathbf{k}} &=&(a\psi _{1\mathbf{k}}+\psi _{2\mathbf{k}})/\sqrt{
1+a^2}, \\
\beta _{\mathbf{k}} &=&(b\psi _{1\mathbf{k}}+\psi _{2\mathbf{k}})/\sqrt{1+b^2
},
\end{eqnarray*}
where $a$ and $b$ are the functions of $k_x$ and $k_y.$ The mean-field
ground state is obtained by filling all the negative levels and is given by
\[
|\Psi _{\text{mean}}\rangle =\prod_k\beta _{\mathbf{k}}^{\dag }|0\rangle
_\psi .
\]
where the state $|0\rangle _\psi $ is defined through $\psi _{\mathbf{k}
}|0\rangle _\psi =0$. (Note that all the particles $\alpha _{\mathbf{k}}$
has positive energy and all the particles $\beta _{\mathbf{k}}$ has negative
energy.) Since $\beta _{\mathbf{k}}^{\dag }$ is linear combination of $\psi
_1^{\dag }$ and $\psi _2^{\dag }$ and there are $L_x\times L_y$ different $
\mathbf{k}$-levels, the mean-field state $|\Psi _{\text{mean}}\rangle $
contains $L_x\times L_y$ number of fermions. Here $L_{x,y}$ are sizes of the
lattice in the $x$- and $y$-directions.

Clearly, when both $L_x$ and $L_y$ are odd, $|\Psi _{\text{mean}}\rangle $
contains an odd number of fermions. Such a mean-field state does not
correspond to any physical spin state since the corresponding spin wave
function \Eq{Phichieta} vanishes. (Note that \Eq{Phichieta} is a projection
to the subspace with 0 or 2 fermions per site.) To get a non-zero physical
spin wave function we need to start with a mean-field state with one extra
fermion in the empty $\al$-band (or a hole in the filled $\beta $-band). But
by choosing different states for the extra fermion (or the hole), we can
obtain many different spin wave functions which are nearly degenerate. So
when both $L_x$ and $L_y$ are odd, the excitations in the Z2A state are
gapless, or we may say that the Z2A state has infinite degeneracy.
Physically, the Z2A state on odd by odd lattice always contains a unpaired
spinon. The different states of the unpaired spinon gives rise to the
infinite degeneracy.

When one of $L_{x,y}$ is even, the mean-field state $|\Psi _{\text{mean}
}\rangle $ gives rise to a non-zero physical spin state. There is no
unpaired spinon and the excitations are gaped. Each ansatz $u_{ij}^{(m,n)}$
produces a single physical spin state and the Z2A state has four-fold
degeneracy on a torus with an even number of lattice sites.

Because the spin Hamiltonian is translation invariant, the ground states
carry definite crystal momentum. To calculate the crystal momentum, we note
that in the $(m,n)=(0,0)$ sector described by the ansatz $u_{ij}^{(0,0)}$,
the fermion wave function satisfies the periodic boundary condition. So $
(k_x,k_y)$ are quantized as $(k_x,k_y)=(n_x\frac{2\pi }{L_x},n_y\frac{2\pi }{
L_y})$ where $n_{x,y}$ are integers. And the spin state produced by the
ansatz $u_{ij}^{(0,0)}$ has the following crystal momentum:
\begin{align*}
K_x& =\sum k_x=\sum_{n_x=1}^{L_x}\sum_{n_y=1}^{L_y}n_x\frac{2\pi }{L_x}=
\frac{L_yL_x(L_x+1)}2\frac{2\pi }{L_x}, \\
K_y& =\sum k_y=\sum_{n_x=1}^{L_x}\sum_{n_y=1}^{L_y}n_y\frac{2\pi }{L_y}=
\frac{L_xL_y(L_y+1)}2\frac{2\pi }{L_y}.
\end{align*}
We would like to point out that the above crystal momentum is actually the
crystal momentum of the mean-field state. However, the even-fermion-per-site
projection commutes with the translation operator, and thus the crystal
momentum is unchanged by projection.

When $m$ and/or $n$ are equal to 1, the fermion wave function is
antiperiodic in the $y$- and/or $x$-directions. In the case, $k_y$ and/or $
k_x$ are quantized as $(n_y+\frac 12)\frac{2\pi }{L_y}$ and/or $(n_x+\frac
12)\frac{2\pi }{L_x}$. The crystal momentum of the spin state produce by the
ansatz $u_{ij}^{(m,n)}$ can be calculated in the similar fashion. For
example in the $(m,n)=(1,1)$ sector, the crystal momentum is given by
\begin{align*}
K_x& =\sum_{n_x=1}^{L_x}\sum_{n_y=1}^{L_y}(n_x+\frac 12)\frac{2\pi }{L_x}=
\frac{L_yL_x(L_x+2)}2\frac{2\pi }{L_x}, \\
K_y& =\sum_{n_x=1}^{L_x}\sum_{n_y=1}^{L_y}(n_y+\frac 12)\frac{2\pi }{L_y}=
\frac{L_xL_y(L_y+2)}2\frac{2\pi }{L_y}.
\end{align*}
The results are summarized in the table \ref{Z2Acm}.

\begin{table}[t]
\begin{tabular}{|c|cccc|}
\hline
$(K_x,K_y)$ & (ee) & (eo) & (oe) & (oo) \\ \hline
$(00)$ & $(0,0)$ & $(\pi,0)$ & $(0,\pi)$ & -- \\
$(01)$ & $(0,0)$ & $(\pi,0)$ & $(0,0)$ & -- \\
$(10)$ & $(0,0)$ & $(0,0)$ & $(0,\pi)$ & -- \\
$(11)$ & $(0,0)$ & $(0,0)$ & $(0,0)$ & -- \\ \hline
\end{tabular}
\caption{Crystal momenta $(K_x,K_y)$ of the four ground states, $(m,n)$=
(0,0), (0,1), (1,0), (1,1), of the Z2A spin liquid on three different
lattices, $(L_x,L_y)=$ (even,even), (even,odd), (odd,even).}
\label{Z2Acm}
\end{table}

\section{Topological properties for the exact soluble model}

To understand the topological order in the Z2E state of the exact soluble
model, we would like to calculate the ground state degeneracy and ground
state crystal momenta of the Z2E state. Just like the Z2A state discussed in
the last section, one can construct many-spin wave functions of the
degenerate ground states from the mean-field ansatz
%
Eq. (\ref{sl5.1}) 
with $(u_{ij},\eta _{ij})$ given by Eq. (\ref{Z2Eans}). The four
mean-field ansatz $(u_{ij}^{(m,n)},\eta _{ij}^{(m,n)})$ can
potentially give rise to four degenerate ground states. But some
time, the mean-field ground state contains odd numbers of fermions.
In this case, the corresponding mean-field ansatz does not lead to
physical spin wave function.


To calculate the fermion number in the mean-field ground state, one can
write down the ``mean-field'' fermion Hamiltonian in momentum space
\begin{widetext}
\begin{eqnarray}
H_\text{mean}(\mathbf{k}) &=&\sum_{\v k>0}(\psi
_{1\mathbf{k}}^{\dagger },\psi _{1,-\mathbf{k}})(
\begin{array}{ll}
\cos k_x & \imth \sin k_x \\
-\imth \sin k_x & -\cos k_x
\end{array}
)\left(
\begin{array}{l}
\psi _{1\mathbf{k}} \\
\psi _{1,-\mathbf{k}}^{\dagger }
\end{array}
\right) +\sum_{\v k>0}(\psi _{2\mathbf{k}}^{\dagger },\psi _{2,-\mathbf{
k}})(
\begin{array}{ll}
\cos k_y & \imth \sin k_y \\
-\imth \sin k_y & -\cos k_y
\end{array}
)\left(
\begin{array}{l}
\psi _{2\mathbf{k}} \\
\psi _{2,-\mathbf{k}}^{\dagger }
\end{array}
\right)  \nonumber  \label{h2} \\
&&+\psi _{1\mathbf{k}}^{\dag }\psi _{1\mathbf{k}}\Big|_{k_x=0,k_y=0}-\psi _{1\mathbf{k}
}^{\dag }\psi _{1\mathbf{k}}\Big| _{k_x=\pi ,k_y=\pi }+\psi
_{2\mathbf{k}}^{\dag
}\psi _{2\mathbf{k}}\Big|_{k_x=0,k_y=0}-\psi _{2\mathbf{k}}^{\dag }\psi _{2\mathbf{k}
}\Big| _{k_x=\pi ,k_y=\pi }  \nonumber \\
&& +\psi _{1\mathbf{k}}^{\dag }\psi
_{1\mathbf{k}}\Big|_{k_x=0,k_y=\pi } -\psi _{1\mathbf{k} }^{\dag
}\psi _{1\mathbf{k}}\Big| _{k_x=\pi ,k_y=0} -\psi
_{2\mathbf{k}}^{\dag }\psi _{2\mathbf{k}}\Big|_{k_x=0,k_y=\pi }
+\psi _{2\mathbf{k}}^{\dag }\psi _{2\mathbf{k}}\Big| _{k_x=\pi
,k_y=0}
\nonumber \\
&=&\sum_{\v k>0} [ \alpha _{\mathbf{k}}^{\dag }\alpha _{\mathbf{k} }
+\alpha _{-\mathbf{k}}^{\dag }\alpha _{-\mathbf{k} } ] + \sum_{\v
k>0}[ \beta _{\mathbf{k}}^{\dag }\beta _{\mathbf{k} } +\beta
_{-\mathbf{k}}^{\dag }\beta _{-\mathbf{k} } ]
\nonumber \\
&&+\psi _{1\mathbf{k}}^{\dag }\psi _{1\mathbf{k}}\Big|_{k_x=0,k_y=0}-\psi _{1\mathbf{k}
}^{\dag }\psi _{1\mathbf{k}}\Big| _{k_x=\pi ,k_y=\pi }+\psi
_{2\mathbf{k}}^{\dag
}\psi _{2\mathbf{k}}\Big|_{k_x=0,k_y=0}-\psi _{2\mathbf{k}}^{\dag }\psi _{2\mathbf{k}
}\Big| _{k_x=\pi ,k_y=\pi }  \nonumber \\
&& +\psi _{1\mathbf{k}}^{\dag }\psi
_{1\mathbf{k}}\Big|_{k_x=0,k_y=\pi } -\psi _{1\mathbf{k} }^{\dag
}\psi _{1\mathbf{k}}\Big| _{k_x=\pi ,k_y=0} -\psi
_{2\mathbf{k}}^{\dag }\psi _{2\mathbf{k}}\Big|_{k_x=0,k_y=\pi }
+\psi _{2\mathbf{k}}^{\dag }\psi _{2\mathbf{k}}\Big|_{k_x=\pi
,k_y=0},
\end{eqnarray}
\end{widetext}
with
\begin{eqnarray*}
\left(
\begin{array}{l}
\alpha _{\mathbf{k}} \\
\alpha _{-\mathbf{k}}^{\dagger }
\end{array}
\right) &=&\exp (-\imth k_x\Big(
\begin{array}{ll}
0 & 1 \\
1 & 0
\end{array}
\Big))\left(
\begin{array}{l}
\psi _{1\mathbf{k}} \\
\psi _{1,-\mathbf{k}}^{\dag }
\end{array}
\right) , \\
\left(
\begin{array}{l}
\beta _{\mathbf{k}} \\
\beta _{-\mathbf{k}}^{\dagger }
\end{array}
\right) &=&\exp (-\imth k_y\Big(
\begin{array}{ll}
0 & 1 \\
1 & 0
\end{array}
\Big))\left(
\begin{array}{l}
\psi _{2\mathbf{k}} \\
\psi _{2,-\mathbf{k}}^{\dag }
\end{array}
\right) .
\end{eqnarray*}
Here $\mathbf{k}=0$ means that $(k_x,k_y)=(0,0)$, $(0,\pi )$,
$(\pi ,0)$, or $(\pi ,\pi )$, and $\mathbf{k}>0$ means that $k_y>0$ or
$k_y=0,\ k_x>0$ \emph{and} $\v k\neq 0$.

We note that both $\alpha $ band and $\beta $ band have a positive energy $
E_{\mathbf{k}}=1$. $\alpha _{\pm \mathbf{k}}$, $\beta _{\pm \mathbf{k}}$
will annihilate the mean-field ground state $|\Psi _{\text{mean}}\rangle ,$
\[
\alpha _{\pm \mathbf{k}}|\Psi _{\text{mean}}\rangle =0,\ \ \ \ \ \ \ \ \beta
_{\pm \mathbf{k}}|\Psi _{\text{mean}}\rangle =0.
\]
It needs to point out that the above formula for the ``mean-field'' fermion
Hamiltonian are valid only for even-by-even lattice with periodic boundary
condition, i.e. $(m,n)=(0,0)$. For other cases (even-by-even lattice with
anti-periodic boundary conditions, and even-by-odd, odd-by even, odd by odd
lattices with both periodic boundary condition and anti-periodic boundary
conditions), one or more of the four high symmetry points at momentum space $
\mathbf{k}^{*}=(0,0)$, $(0,\pi )$, $(\pi ,0)$, $(\pi ,\pi )$ are absent
which is shown in the table in appendix.

We also note that, for $\mathbf{k}\neq 0$,
\begin{align*}
\al_{\v k} &=      u_{\v k} \psi_{1,\v k} +   v_{\v k} \psi^\dag_{1,-\v k}
\nonumber\\
\al^\dag_{-\v k}&=-v^*_{\v k} \psi_{1,\v k} + u^*_{\v k} \psi^\dag_{1,-\v k} .
\end{align*}
The condition
$\al_{\v k}|\Phi_\text{mean}\> = \al_{-\v k}|\Phi_\text{mean}\>=0 $
implies that (if we only consider the $\v k$ and $-\v k$ levels)
\begin{equation*}
 |\Phi_\text{mean}\>=
(v_{\v k}  + u_{\v k}\psi^\dag_{1,-\v k}\psi^\dag_{1,\v k})
|0\>
\end{equation*}
We see that $\mathbf{k}\neq 0$ levels always contribute even numbers of
fermions.  Also, since $v_{\v k}  + u_{\v k}\psi^\dag_{1,-\v k}\psi^\dag_{1,\v
k}$ carries $0$ momentum, we see that the contribution to the total momentum
from the $\mathbf{k}\neq 0$ levels is zero.

%

Thus to determine if the mean-field ground state contain even or odd number
of $\psi $ fermions, we only need to examine the occupation on the
four $\v k=0$ momentum points: $\mathbf{k}=(0,0)$, $(0,\pi )$, $(\pi ,0)$,
$(\pi ,\pi)$. The Hamiltonian on those four points is contained in
Eq. (\ref{h2}).
All the negative energy levels are filled in the mean-field ground
state.
On an even by even lattice and for the $(m,n)=(0,0)$ ansatz,
all the momenta $(\pi ,0)$, $(0,\pi )$, and $(\pi ,\pi
) $ are allowed. Thus the $(\pi ,0)$ level and the $(\pi ,\pi )$ level
each is occupied by a $\psi _1$ fermion, and the $(0,\pi )$ level and the
$(\pi ,\pi )$ level each is occupied by a
$\psi _2$ fermion. The total momentum of the ground state is $(\pi ,\pi )$.
Such a mean-field ground state has even numbers of fermions. It will survive
the projection and lead to a physical spin ground state.
Other situations can be calculated in the same way. Here we only
summarize the result: on an even by even lattice, there exist four different
degenerate ground states. However, on other kinds of lattice ( even by odd,
odd by even and odd by odd), there exist only two different ground states.
The other two states are projected out since the mean-field ground states
contain odd numbers of fermions. The crystal momenta of the degenerate
ground states can also be calculated which are summarized in table \ref{Z2Ecm}.

\begin{table}[t]
\begin{tabular}{|c|cccc|}
\hline
$(K_x,K_y)$ & (ee) & (eo) & (oe) & (oo) \\ \hline
$(00)$ & $(\pi,\pi)$ & -- & -- & -- \\
$(01)$ & $(0,0)$ & -- & $(0,0)$ & $(0,0)$ \\
$(10)$ & $(0,0)$ & $(0,0)$ & -- & $(0,0)$ \\
$(11)$ & $(0,0)$ & $(0,0)$ & $(0,0)$ & -- \\ \hline
\end{tabular}
\caption{Crystal momenta of the degenerate ground states, $(m,n)$= (0,0),
(0,1), (1,0), (1,1), of the Z2E spin liquid on four different lattices, $
(L_x,L_y)=$ (even,even), (even,odd), (odd,even), (odd,odd).}
\label{Z2Ecm}
\end{table}

\section{The mutual \textrm{U(1)}$\times $\textrm{U(1)} CS theory}

In the above sections we have calculated the topological properties for the
Z2A and the Z2E states. Due to their different topological properties, we
find that the two states have different topological orders. Then an
important issue is to find the low energy effective theories that describe
the two different topological orders. We find that a mutual \textrm{U(1)}$
\times $ \textrm{U(1)} CS theory with different projective realizations of
the lattice symmetry can describe the two kind of topological orders. We
reach the conclusion by comparing the topological properties of the mutual
\textrm{U(1)}$\times $\textrm{U(1)} CS theory with those of the Z2A and the
Z2E states. All the topological properties, including topological
degeneracy, quantum numbers, and edge states, agree, indicating the
equivalence between the $Z_2$ topological states on lattice and the mutual
\textrm{U(1)}$\times $\textrm{U(1)} CS theory.

\subsubsection{Mutual \textrm{U(1)}$\times $\textrm{U(1)} CS theory}

First we introduce the Lagrangian for the mutual \textrm{U(1)}$\times $
\textrm{U(1)} CS theory :
\begin{eqnarray}
\label{mCSL}
\mathcal{L}_{\mathrm{eff}} &=&-\frac 1{4e_a^2}(f_{\mu \nu })^2-\frac
1{4e_A^2}(F_{\mu \nu })^2  \label{mcs} \\
&&+\frac 1\pi \epsilon ^{\mu \nu \lambda }A_\mu \partial _\nu a_\lambda +\imth
a^\mu j_\mu +\imth A^\mu J_\mu
\end{eqnarray}
where $f_{\mu \nu }$ is the gauge field strength for gauge field $ a_\lambda $
and $F_{\mu \nu }$ is the gauge field strength for gauge field $ A_\mu $.  The
excitations are described by the currents which are defined as $j_\mu
=(j_i,\rho _a)$ and $J_\mu =(J_i,\rho _A)$.  The gauge charges of $a_\mu$ and
$A_\mu$ are quantized as integers.

%
From the equation motions for $a_\lambda $ and $A_\lambda ,$
\begin{eqnarray*}
-\frac 1{2e_a}(\partial _\mu f_{\mu \lambda })+\frac 1\pi \epsilon ^{\mu \nu
\lambda }F_{\mu \nu } &=&-\imth j_\mu , \\
-\frac 1{2e_A^2}(\partial _\mu F_{\mu \lambda })+\frac 1\pi \epsilon ^{\mu
\nu \lambda }f_{\mu \nu } &=&-\imth J_\mu ,
\end{eqnarray*}
we find that a \textrm{U(1)} charge for gauge field $A_\mu $ induces flux of
gauge field $a_\mu $. As a result, the \textrm{U(1)} charge for gauge field $
A_\mu $ and the \textrm{U(1)} charge for gauge field $a_\mu $ have a
semionic mutual statistics. That is, moving an $A_\mu $-charge around an $
a_\mu $-charge generates a phase $\pi $. This catches the key topological
property for the $Z_2$ spin liquid. It is well known that the $Z_2$ spin
liquid states contain $Z_2$ vortex and $Z_2$ charge excitations. And the $
Z_2 $ vortex and the $Z_2$ charge have semionic mutual statistics between
them.
So we will propose that the mutual Chern-Simons theory
\ref{mCSL} describes a $Z_2$ gauge theory.
The $A_\mu $-charge can be identified as the $Z_2$ charge and the
$a_\mu $-charge as the $Z_2$ vortex. 

Furthermore, the energy gap for both of the gauge fields comes from the
mutual CS term
\[
m_a\sim e_a^2,\ \ \ \ \ \ m_A\sim e_A^2.
\]
The mutual \textrm{U(1)}$\times $\textrm{U(1)} CS theory describes a gapped
topological state. This also agrees with the $ Z_2 $ topological states where
all excitations are gapped.

However, we have two kinds of $Z_2$ topological orders Z2A and Z2E. How can
the two different $Z_2$ topological orders be described by the same \textrm{
U(1)}$\times $\textrm{U(1)} CS theory? In the following we will show that two
different $Z_2$ topological orders are described by the same \textrm{U(1)
}$\times $\textrm{U(1)} CS theory but with different realizations of the
lattice symmetry.

To obtain two different realizations of lattice symmetry,
we note that $Z_2$ vortices for the exactly soluble model (the Z2E state)
live on the even plaquettes. The vortices on the odd plaquettes are
actually the $Z_2$ charge.\cite{Wqoexct,Wen04} So under a translation by one
lattice spacing, a $Z_2$ vortex is changed into a $Z_2$ charge! So in the
mutual \textrm{U(1)}$\times $\textrm{U(1)} CS theory that describes the Z2E
state, $a_\mu $ and $A_i$ must exchange under the translation by one lattice
spacing.

Also, the Z2A state contains $\pi$ flux through each square.  This
$\pi$ flux also affects how $a_\mu$ is transformed under
translation.  To see this, let us consider two Wilson loop operators
$W_1=e^{\imth \oint_{C_1} dy a_y}$ and $W_2=e^{\imth \oint_{C_2} dy
a_y}$ along two loops $C_1$ and $C_2$.  Both loops wrap around the
torus in $y$-direction. However the loop $C_2$ is displaced from the
loop $C_1$ by one lattice constant in the $x$-direction. In the
following, we will assume the lattice constant is $a=1$.  Due to the
$\pi$ flux through each square, we see that $W_2=(-)^{L_y} W_1$,
where $L_y$ is the length of the torus in the $y$-direction. So
under a translation by one lattice constant in the $x$-direction,
$a_y$ must change to $a_y+\pi$, to account for the change in the
Wilson loop.

The above discussion motivates us to define two types of mutual \textrm{U(1)}
$\times $\textrm{U(1)} CS theories which have different realizations of
translation symmetries. Let ${T}_x$ and ${T}_y$ be the translations
by one lattice spacing in the $x$ and $y$ directions respectively.
%
The first type of the mutual \textrm{U(1)}$\times $\textrm{U(1)} CS theory
is denoted as Z2A type which describes the Z2A state.
%
The $\pi$ flux makes the
gauge fields transform non-trivially under translations:
\begin{align}
T_x^{-1}  A_xT_x &=  A_x,& T_y^{-1}  A _xT_y&=  A_x +\pi,
\nonumber  \label{AaTrans} \\
T_x^{-1}  A_yT_x &=  A_y+\pi ,& T_y^{-1}  A _yT_y&=  A_y,
\nonumber \\
T_x^{-1}  a_xT_x &=  a_x,& T_y^{-1}  a _xT_y&=  a_x+\pi ,
\nonumber \\
T_x^{-1}  a_yT_x &=  a_y +\pi,& T_y^{-1}  a _yT_y&=  a_y.
\end{align}
Since the translation $T_x$ ($T_y$) may shift $ A_y$ ($a_x)$ by $\pi $,
this reproduces the different patterns of crystal momenta of the degenerate
ground states on different lattices.

The other type of the mutual CS theory is denoted as Z2E type that describes
the Z2E state. It has no flux.
However, the gauge fields still transform non-trivially under
translations:
\[
T_i^{-1} A_jT_i= a_j,\ \ \ \ \ \ T_i^{-1} a_jT_i= A_j,\
\ \ \ \ \ i=x,y
\]
$ A_i$ and $ a_i$ will exchange under a translation operation by
one lattice spacing.

\subsubsection{The topological degeneracy}

In the next a few sections, we will calculate the topological properties of
the above two types of mutual CS theory. First, we calculate the topological
degeneracy for the ground states.
In the temporal gauge, $A_0=0,$ and on an even-by-even lattice, the
fluctuations $ A_i$ and $ a_i$ are periodic. We can expand them as
\begin{align}
( A_x, A_y)& =(\frac 1{L_x}\Theta _x+\sum_{\mathbf{k}}A_{\mathbf{k}
}^xe^{\imth \check{x}\cdot \mathbf{k}},\frac 1{L_y}\Theta _y+\sum_{\mathbf{k}}A_{
\mathbf{k}}^ye^{\imth \check{x}\cdot \mathbf{k}}),  \label{modeAa} \\
( a_x, a_y)& =(\frac 1{L_x}\theta _x+\sum_{\mathbf{k}}a_{\mathbf{k}
}^xe^{\imth \check{x}\cdot \mathbf{k}},\frac 1{L_y}\theta _y+\sum_{\mathbf{k}}a_{
\mathbf{k}}^ye^{\imth \check{x}\cdot \mathbf{k}})
\end{align}
where $\mathbf{k=}(k_x,k_y)=(\frac{2\pi }{L_x}n_x,$ $\frac{2\pi }{L_y}n_y)$
where $n_{x,y}$ are integers. $(A_{\mathbf{k}}^x,A_{\mathbf{k}}^y)$ and $(a_{
\mathbf{k}}^x,a_{\mathbf{k}}^y)$ are the gauge fields with non-zero momentum
and $\left( \Theta _x,\Theta _y\right) $ and $\left( \theta _x,\theta
_y\right) $ are the zero modes with zero momentum for the gauge fields $A_i$
and $a_i$. Because the existence of the mass gap, the degree freedoms for
gauge fields with non-zero momentum $\left( A_{\mathbf{k}}^x,A_{\mathbf{k}
}^y\right) \ $and $\left( a_{\mathbf{k}}^x,a_{\mathbf{k}}^y\right) $ have
nothing to do with the low energy physics. It is the degree freedoms of zero
momentum $\left( \Theta _x,\Theta _y\right) $ and $\left( \theta _x,\theta
_y\right) $ that determine the low energy physics. The effective Lagrangian
Eq.(\ref{mcs}) determines the dynamics of $\left( \Theta _x,\Theta _y\right)
$ and $\left( \theta _x,\theta _y\right) $ which corresponds to two
particles on a plane with a finite magnetic field. $(\Theta _x,\theta _y)$
are the coordinates of the first particle, and $(\Theta _y,\theta _x)$ are
the coordinates of the second particle. Thus we map the original mutual
\textrm{U(1)}$\times $\textrm{U(1)} CS theory to a quantum mechanics model
of two particles (see appendix). The energy spectrum for the quantum
mechanics model can be solved easily. The lowest energy levels for above
model reveal the topological characters for the ground states. The degeneracy
for $(\Theta _x,\theta _y)$ degrees of freedom and the degeneracy for $
(\Theta _y,\theta _x)$ degrees of freedom are given as $D_{(\Theta _x,\theta
_y)}=2$ and $D_{(\Theta _y,\theta _x)}=2$. For both the Z2A type and the Z2E
type CS theories, there exist four degenerate ground states
\[
D=D_{(\Theta _x,\theta _y)}D_{(\Theta _y,\theta _x)}=2\times 2=4.
\]

However, the above result only applies to even-by-even lattice. For other
cases, even-by-odd, odd-by-even and odd-by-odd, the situations are changed.
We will discuss those more complicated cases in appendix.. We find that for
the Z2A type mutual CS theory, the ground state degeneracy remain to be $4$
for even-by-odd and odd-by-even lattices. For the Z2E type mutual CS theory,
the ground state degeneracy becomes $2$ for even-by-odd, odd-by-even, and
odd-by-odd lattices.

One way to understand the later result is to note that if $L_x$ is odd then
one gauge field will turn into the other one as we go around the lattice
along $x$-direction. Thus the gauge fields have a twisted boundary
condition:
\[
 A_i(x+L_x,y)= a_i(x,y),\ \ \ \ \ \  a
_i(x+L_x,y)= A_i(x,y).
\]
This twisted boundary condition means that $ A_\mu $ and $ a_\mu $
can be viewed as a single gauge field on a lattice whose size is doubled in
the $x$-direction. There are only two zero modes in the mode expansion. As a
result the ground-state degeneracy on even-by-odd, odd-by-even and
odd-by-odd is reduced to $2$. We can also use the CS theories to calculate
the crystal momenta of the ground states (see appendix). The results agree
with those in tables \ref{Z2Acm} and \ref{Z2Ecm}.

\subsubsection{The edge states}

We can also use the mutual \textrm{U(1)}$\times $\textrm{U(1)} CS theories
to study edge excitations. First, let us consider the exact soluble model (
\ref{Hexct}) on a finite $L_x\times L_y$ lattice with a periodic boundary
condition only along $y$-direction. The lattice has two edges along $y$
-direction located at $i_x=0$ and $i_x=L_x$. Such a lattice model can be
obtained from the periodic lattice model (\ref{Hexct}) by setting $g=0$ for
a column of plaquettes. The resulting model is still exactly soluble. We
find that the ground states have $\sim 2^{L_y}$-fold degeneracy which arise
from $\sigma _i^y\sigma _{i+\check{x}}^x\sigma _{i+\check{x}+\check{y}
}^y\sigma _{i+\check{y}}^x=\pm 1$ on the column of plaquettes with $g=0$.
Those degenerate states can be viewed as gapless edge excitations on the two
boundaries. Since there are $2L_y$ edge sites, we find that there are $\sqrt{
2}$ edge states per edge site, indicating that the gapless edge states are
described by Majorana fermions. Indeed, the gapless edge excitations can be
mapped to a Majorana fermion system exactly. 

To obtain the gapless edge states from the mutual CS theories, we introduce
\[
a_{+,\mu }=A_\mu +a_\mu ,\ \ \ \ \ \ \ \ \ \ \ \ a_{-,\mu }=A_\mu -a_\mu
\]
and rewrite the mutual $U(1)\times U(1)$ CS effective theory as
\begin{equation}
\cL_{\text{eff}}=\frac 1{4\pi }a_{+,\mu }\prt_\nu a_{+,\la}\eps^{\mu \nu \la
}-\frac 1{4\pi }a_{-,\mu }\prt_\nu a_{-,\la}\eps^{\mu \nu \la}+...
\label{CSpm}
\end{equation}
The charges of $A_\mu $ and $a_\mu $ are quantized as integers. Converting
the $A_\mu $ and $a_\mu $ charges to the $a_{+,\mu }$ and $a_{-,\mu }$
charges, we find that the $a_{+,\mu }$ and $a_{-,\mu }$ charges are still
quantized as integers. However, $(1/2,1/2)$ charge for the $a_{+,\mu }$ and $
a_{-,\mu }$ field is also allowed.

The mutual CS theory (\ref{CSpm}) has one right-moving and one left-moving
branches of edge excitations. The two branches of the edge excitations are
described by the following 1D fermion theory \cite{Wen04}
\[
\cL_{\text{edge}}=\psi _R^{\dag }(\prt_t-v\prt_x)\psi _R+\psi _L^{\dag }(\prt
_t+v\prt_x)\psi _L+...
\]
at low energies, where $(...)$ represent terms that are consistent with the
underlying symmetries of the lattice model. $\psi _R$ carries a unit of $
a_{+}$ charge and $\psi _{-}$ a unit of $a_{-}$ charge. We note that the $
A_\mu $ and $a_\mu $ charges, as the $Z_2$ charge and the $Z_2$ vortex, are
conserved only mod 2. So $(...)$ may contain terms that change $
(a_{+},a_{-}) $ charge by $(1,1)$ and $(1,-1)$. Thus, the following terms
\[
a\psi _R\psi _L+b\psi _R\psi _L^{\dag }+h.c.
\]
are allowed in the low energy effective Lagrangian. The additional terms
will open an energy gap for the edge excitations and one may conclude that
the Z2E state in the exactly soluble model (\ref{Hexct}) has no gapless edge
excitations in general. 

However, the above conclusion is not quite correct. We see that although the
presence of the edge breaks the translation symmetry in the $x$-direction,
the finite system still has the translation symmetry in the $y$-direction.
Under the translation in the $y$-direction by lattice spacing, $A_\mu $ and $
a_\mu $ is exchanged, or $(a_{+,\mu },$ $a_{-,\mu })$ are changed into $
(a_{+,\mu },$ $-a_{-,\mu })$. So the translation in the $y$-direction
changes the sign of the $a_{-}$ charge and hence changes $\psi _L$ to $\psi
_L^{\dag }$. As a result, only the following term
\[
a\psi _R(\psi _L+\psi _L^{\dag })+h.c.
\]
can be added to the edge effective Lagrangian, which do not break the
translation symmetry along the edge.

Introducing Majorana fermions
\[
\psi _R=\la_R+\imth \eta _R,\ \ \ \ \ \ \ \psi _L=\la_L+\imth \eta _L,
\]
we can rewrite the edge effective Lagrangian as
\begin{align*}
\cL_{\text{edge}}& =\la_R(\prt_t-v\prt_x)\la_R+\eta _R(\prt_t-v\prt_x)\eta _R
\\
& \ \ \ +\la_L(\prt_t+v\prt_x)\la_L+\eta _L(\prt_t+v\prt_x)\eta _L \\
& \ \ \ +2(a\la_R\la_L+\imth a\la_R\eta _L+h.c.).
\end{align*}
The $a\la_R\la_L+\imth a\la_R\eta _L$ term gaps a pair of Majorana fermions and
leave the other pair gapless. So the Z2E state has right-moving and
left-moving gapless edge excitations described by Majorana fermions,
provided that the edge is in the $x$- or $y$-direction. The presence of the
translation symmetry in the $x$- or $y$-direction is crucial for the
existence of the gapless edge excitations for the Z2E type mutual \textrm{
U(1)}$\times $\textrm{U(1) }CS theory and the exact soluble model.

For the Z2A state, although the low energy effective theory has the same
form as the exactly soluble model, the translation does not induce the
exchange between $A_\mu $ and $a_\mu $. As a result, in general, there are
no gapless edge excitations for the Z2A type mutual \textrm{U(1)}$\times $
\textrm{U(1) }CS theory and the Z2A state.

\section{Conclusion}

In this paper, two kinds of $Z_2$ topological ordered states for frustrated
spin systems, Z2A state and Z2E state, are studied. Using the \textrm{SU(2)}
slave-particle theory, we calculate their ground-state degeneracy, their
ground-state quantum numbers, their gapless edge state, and the projective
symmetry group of their quasi-particles. We propose a mutual \textrm{U(1)}$
\times $\textrm{U(1) } Chern-Simons theory with two different realizations
of lattice symmetry as the effective field theories that describe the two
types of topological orders. We show that the effective theories produce the
same low energy physics, including the degeneracy of the ground state, the
quantum number for the ground state and the edge states. It turns out that
the different $Z_2$ topological orders are reflected in different
realizations of the lattice symmetry in the same effective mutual
Chern-Simons theory.

We like to mention that the Z2A phase appears to be an example of
``weak symmetry breaking in dimension 2'', while the Z2E phase appears to be an
example of ``weak symmetry breaking in dimension 1'' discussed in Ref.
\onlinecite{K062}. So these two phases are
examples of the two basic ways that lattice symmetries and topological
structure can be entangled.

This research is supported by NSF Grant No. DMR-0706078,
NFSC no. 10228408, and NFSC no. 10574014.

\appendix

\section{Appendix}

\subsubsection{Topological degeneracy for the Z2E state}

We have used ansatz $(u_{ij}^{(m,n)},\eta
_{ij}^{(m,n)})=((-)^{ms_x(ij)}(-)^{ns_y(ij)}\bar{u}
_{ij},(-)^{ms_x(ij)}(-)^{ns_y(ij)}\bar{\eta}_{ij})$ to describe the
four degenerate ground states for the Z2E state. Here $m,n=0,1$.
$s_{x,y}(ij)$ have values $0$ or $1$, with $s_{x,y}(ij)=1$ if the
link $ij$ crosses the $x$ or $y$ line (see Fig.~\ref{xyline}) and
$s_{x,y}(ij)=0$ otherwise.

It is pointed out that the above result of four degenerate ground states is
right only for the Z2E state on an even-by-even lattice. On other kinds of
lattice ( even by odd, odd by even and odd by odd), there exist only two
different ground states. The other two states are projected out since the
mean-field ground states contain odd numbers of fermions.

Let's calculate the topological degeneracy for the Z2E state on different
lattices in detail. It was pointed out that the total number of the $\psi $
fermions on $\mathbf{k}$ and $-\mathbf{k}$ is always even if $\mathbf{k}\neq
0$. To determine if the mean-field ground state contains even or odd number
of $\psi $ fermions, we will only pay attention to the occupation on the
following four momentum points: $\mathbf{k}=(0,0)$, $(0,\pi )$, $(\pi ,0)$,
$(\pi ,\pi )$.

Firstly, we discuss the topological degeneracy for Z2E state on an even by
even lattice.  For the ground state described by $(m,n)=(0,0)$, the energy
levels for both $\psi_1$ and $\psi_2$ have positive energies at $\v k=(0,0)$
(see Eq. \ref{h2}).  Thus the $\mathbf{k}=(0,0)$ level is not  occupied.  We
also see from Eq. \ref{h2} that, at  $\v k=(0,\pi)$, $\psi_1$ has a  positive
energy $\psi_2$ has a negative energy.  Thus the $\mathbf{k}=(0,\pi)$ level is
occupied by a $\psi _2$ particle. Similarly, we find that the
$\mathbf{k}=(\pi,0)$ level is occupied by a $\psi _1$ particle, the
$\mathbf{k}=(\pi,\pi)$ level is occupied by a $\psi _1$ particle and a $\psi
_2$ particle.  Therefore, four particles occupy the points $(0,0)$, $(0,\pi
)$, $(\pi ,0)$, $(\pi ,\pi )$. Because the meanfield ground state $|\Psi
_{\text{mean}}^{(u_{ij}^{(0,0)},\eta _{ij}^{(0,0)})}\rangle $ has even number
particles, it survives the even-particle-per-site projection.

Also, the total contribution to the crystal momentum from the $\v k\neq 0$
levels is zero.  Thus the total crystal momentum is determined by the
particles that occupy the $(0,0)$, $(0,\pi)$, $(\pi ,0)$, $(\pi ,\pi )$
levels. We find that the total crystal momentum
of the above state is $0\times (0,0) + 1\times (0,\pi) +1\times (\pi ,0)
+2\times (\pi ,\pi )=(\pi ,\pi )$.


For the ground states described by $(m,n)=(1,0),$ $(m,n)=(0,1)$, and
$(m,n)=(1,1)$, non of the high-symmetry points $(0,0)$, $(0,\pi)$, $(\pi ,0)$,
$(\pi ,\pi )$ exist. Thus the ground states have even number particles, so
they are all permitted under the even-particle-per-site projection.  The total
crystal momenta of the above states are all zero.

Therefore, there are four degenerate ground states on even-by-even lattice.
One carries crystal momentum $(\pi,\pi)$ and other three carry crystal
momentum $(0,0)$.  This corresponds to the first column of table \ref{Z2Ecm}.


Secondly, we discuss the topological degeneracy for Z2E state on an even by
odd lattice.  For the ground state described by $(m,n)=(0,0)$, the $\v
k=(0,0)$ level is not occupied; the $\v k=(\pi ,0)$ level is occupied by one
$\psi _1$ particle, as before.  The points $(0,\pi )$ and $(\pi ,\pi )$ do not
exist. As a result, only one particle occupies the high-symmetry points.
Because the ground state $ |\Psi _{\text{mean}}^{(u_{ij}^{(0,0)},\eta
_{ij}^{(0,0)})}\rangle $ has odd number particles, it is forbidden by the
even-particle-per-site projection.


For the ground state described by $(m,n)=(0,1)$, the points $(0,0)$, $(0,\pi
)$, $(\pi ,0)$, $(\pi ,\pi )$ do not exist. Thus the ground state has even
number particles, so it is permitted by the projection.
Such a state carries a $(0,0)$ crystal momentum.


For the ground state described by $(m,n)=(1,0)$, the $\v k=(0,\pi )$ level is
occupied by a $\psi _2$ particle, and the $\v k=(\pi ,\pi )$ level is occupied
by a $\psi _1$ and a $\psi _2$ particles.  The $(\pi ,0)$ and $(0,0)$ points
do not exist. As a result, three particles occupy
the high-symmetry points.
The state is forbidden by the projection.


For the ground state noted by $(m,n)=(1,1)$, the points $ (0,0)$, $(0,\pi )$,
$(\pi ,0)$, $(\pi ,\pi )$ do not exist. Because the ground state $|\Psi
_{\text{mean}}^{(u_{ij}^{(1,1)},\eta _{ij}^{(1,1)})}\rangle $ has even number
particles, it is also permitted by the projection.  Such a state also carries
a $(0,0)$ crystal momentum.


Therefore there are two degenerate ground states on an even by odd
lattice. Similarly topological degeneracy for Z2E state on an odd by
is also two.  All those states carry a $(0,0)$ crystal momentum.
This corresponds to the second and third columns of table
\ref{Z2Ecm}.

Last, let us discuss the topological degeneracy for Z2E state on an odd by odd
lattice.  For the ground state described by $(m,n)=(0,0)$, the $\v k=(0,0)$
level is not  occupied. The points $(\pi ,0)$ $(0,\pi )$ and $(\pi ,\pi )$ do
not exist. As a result, no particle occupies
the high-symmetry points.
The ground state $|\Psi _{\text{mean}
}^{(u_{ij}^{(0,0)},\eta _{ij}^{(0,0)})}\rangle $ has even number particles
which is permitted by the projection.


For the ground state described by $(m,n)=(1,0)$, the $\v k=(\pi ,0)$ level is
occupied by a $\psi _1$ particle.  The points $(0,0)$ $(0,\pi )$ and $(\pi
,\pi )$ do not exist.  As a result, one particle occupies
the high-symmetry points.
Because the ground state $|\Psi
_{\text{mean}}^{(u_{ij}^{(1,0)},\eta _{ij}^{(1,0)})}\rangle $ has odd number
particles, it is not permitted by the projection.


For the ground state described by $(m,n)=(0,1)$, the $\v k=(0,\pi)$ level is
occupied by a $\psi _2$ particle.  The points $(0,0)$ $(\pi,0 )$ and $(\pi
,\pi )$ do not exist.  As a result, one particle occupies
the high-symmetry points.
Because the ground state $|\Psi
_{\text{mean}}^{(u_{ij}^{(1,0)},\eta _{ij}^{(1,0)})}\rangle $ has odd number
particles, it is not permitted by the projection.


For the ground state described by $(m,n)=(1,1)$, at the $\v k=(\pi ,\pi )$
level is occupied by a $\psi_1$ and a $\psi_2$ particle.  The points $(\pi
,0)$, $(0,\pi )$, and $(0,0)$ do not exist.  As a result, two particle
occupies
the high-symmetry points.
The
ground state $|\Psi _{\text{mean}}^{(u_{ij}^{(1,1)},
\eta_{ij}^{(1,1)})}\rangle $ has even number particles, so the state is
permitted by the projection.


In conclusion, Z2E state has four-fold degeneracy on an
even by even lattice and two-fold degeneracy on an even by odd lattice, odd
by even lattice or odd by odd lattice.
The crystal momenta of those ground states are given by
the table \ref{Z2Ecm}.

\subsubsection{Quantization for the mutual \textrm{U(1)}$\times $\textrm{U(1)
} CS theory}

To calculate the topological properties for the ground states of the mutual
\textrm{U(1)}$\times $\textrm{U(1)} CS theories, one needs to quantize the
gauge fields. We will choose the temporal gauge $A_0=0$. In the temporal
gauge, the physical degrees of freedom are described by $(A_x,A_y)$ and $
(a_x,a_y)$. After the mode expansion, the effective Lagrangian can be
written as
\begin{widetext}
\begin{eqnarray*}
L &=&\frac 12M_x\dot{\Th}_x^2+\frac 12M_y\dot{\Th}_y^2
+\frac 12m_x\dot{\th}_x^2+\frac 12m_y\dot{\th}_y^2
-\frac 1{2\pi }\Th_x\dot{\th}_y-\frac 1{2\pi }\Th_y\dot{\th}_x
+\frac 1{2\pi }\th_y\dot{\Th}_x
+\frac 1{2\pi }\th_x\dot{\Th}_y + ...
\end{eqnarray*}
\end{widetext}
where $(A_{\mathbf{k}}^x,A_{\mathbf{k}}^y)$ and $(a_{\mathbf{k}}^x,a_{
\mathbf{k}}^y)$ represent the terms that certain only the $\mathbf{k}\neq 0$
modes. The masses are given as $M_x=\frac 1{e_A^2}\frac{L_y}{L_x},$ $
M_y=\frac 1{e_A^2}\frac{L_x}{L_y}$ and $m_x=\frac 1{e_a^2}\frac{L_y}{L_x},$ $
m_y=\frac 1{e_a^2}\frac{L_x}{L_y}.$ Because the existence of the mass gap,
the degree freedoms for gauge fields with non-zero momentum $\left( A_{
\mathbf{k}}^x,A_{\mathbf{k}}^y\right) \ $and $\left( a_{\mathbf{k}}^x,a_{
\mathbf{k}}^y\right) $ have nothing to do with the low energy physics. So we
will concentrate on the dynamics of $\theta _{x,y}$ and $\Th_{x,y}$.
\begin{equation}
\mathcal{L}_{\mathrm{eff}}=-\frac 1{4e_a^2}(f_{\mu \nu })^2-\frac
1{4e_A^2}(F_{\mu \nu })^2+\frac 1\pi \epsilon ^{\mu \nu \lambda }A_\mu
\partial _\nu a_\lambda .
\end{equation}

From the effective Lagrangian, one can define the conjugate momentum for
$\left( \Theta _x,\Theta _y\right) $ and $\left( \theta _x,\theta _y\right)$,
\begin{align*}
P_{\Theta _x}&=\frac{\partial L_{\mathrm{eff}}}{\partial \dot{\Theta}_x}=M_x
\dot{\Theta}_x+\frac{\theta _y}{2\pi },
\nonumber\\
P_{\Theta _y}&=\frac{\partial
L_{\mathrm{eff}}}{\partial \dot{\Theta}_y}=M_y\dot{\Theta}_y-\frac{\theta _x
}{2\pi },
\nonumber\\
p_{\theta _x}&=\frac{\partial L_{\mathrm{eff}}}{\partial \dot{\theta}_x}=m_x
\dot{\theta}_x+\frac{\Theta _y}{2\pi },
\nonumber\\
p_{\theta _y}&=\frac{\partial
L_{\mathrm{eff}}}{\partial \dot{\theta}_y}=m_y\dot{\theta}_y-\frac{\Theta _x
}{2\pi }.
\end{align*}

Using the conjugate momentum we write down the following effective
Hamiltonian to describe the low energy physics of the mutual \textrm{U(1)}$
\times $\textrm{U(1)} CS theory
\begin{eqnarray*}
H_{eff} &=&\frac{(P_{\Theta _x}-\frac{\theta _y}{2\pi })^2}{2M_x}+\frac{
(p_{\theta _y}+\frac{\Theta _x}{2\pi })^2}{2m_x} \\
&&+\frac{(P_{\Theta _y}+\frac{\theta _x}{2\pi })^2}{2M_y}+\frac{(p_{\theta
_x}-\frac{\Theta _y}{2\pi })^2}{2m_x}.
\end{eqnarray*}
By choosing different Landau gauges, the above can rewritten as
\begin{equation}
H_{eff}=\frac{(P_{\Theta _x}-\frac{\theta _y}\pi )^2}{2M_x}+\frac{p_{\theta
_y}^2}{2m_y}+\frac{(p_{\theta _x}-\frac{\Theta _y}\pi )^2}{2m_x}+\frac{
P_{\Theta _y}^2}{2M_y}  \nonumber
\end{equation}
or
\begin{equation}
H_{eff}=\frac{P_{\Theta _x}^2}{2M_x}+\frac{(p_{\theta _y}+\frac{\Theta _x}
\pi )^2}{2m_y}+\frac{p_{\theta _x}{}^2}{2m_x}+\frac{(P_{\Theta _y}+\frac{
\theta _x}\pi )^2}{2M_y}.  \nonumber
\end{equation}
The low energy properties of the $U(1)\times U(1)$ Chern-Simons theory
is described by the above Hamiltonian.

\subsubsection{Topological degeneracy and crystal momenta for the Z2E type
mutual \textrm{U(1)}$\times $\textrm{U(1)} CS theory}

Let us first use the Hamiltonian to calculate the ground state degeneracy
of the Z2E state on an even by even lattice.

We note that $a_x$ and $a_x+\frac{2\pi}{L_x}$ are related by a
$U(1)$ gauge transformation. Thus $\th_x=0$ and $\th_x=2\pi$ are
also related by a $U(1)$ gauge transformation, which implies that
$\th_x=0$ and $\th_x=2\pi$ should be viewed as the same point.
Similarly each of the three pairs $\th_y=0$ and $\th_y=2\pi$,
$\Th_x=0$ and $\Th_x=2\pi$, $\Th_y=0$ and $\Th_y=2\pi$, also should
be viewed as the same point.  Thus the above Hamiltonian describes
two particles, each moves on a $2\pi\times 2\pi$ torus.  Each
particle also see $4\pi$ flux through the torus.


The first particle is described by $(\Theta _x,\theta _y)$.
Since there are two units of flux through the torus,
the ground states for the first particle has a degeneracy
$D_{(\Theta _x,\theta _y)}=2$.
Similarly,
the ground states for the second particle also has a degeneracy
$D_{(\Theta _y,\theta _x)}=2$.

As a result, for the Z2E type mutual \textrm{U(1)}$\times $\textrm{U(1)} CS
theory, the ground states have four-fold degeneracy on an even-by-even
lattice :
\begin{equation}
D=D_{(\Theta _x,\theta _y)}D_{(\Theta _y,\theta _x)}=2\times 2=4.  \label{de}
\end{equation}

And the wave-functions $\Psi $ for the four ground states with degenerate
energy are given as $|  1\rangle ,$\ $|  2\rangle ,$\ $|  3\rangle $\
and $|  4\rangle $,
\begin{eqnarray}
\label{wave}
\Psi _1 &\simeq &\exp \left[ -\frac 1{4\pi }\theta _y^2\right] \exp \left[
-\frac 1{4\pi }\Theta _y^2\right] , \\
\Psi _2 &\simeq &e^{-\imth \Theta _x}\exp \left[ -\frac 1{4\pi }\left( \theta
_y-\pi \right) ^2\right] \exp \left[ -\frac 1{4\pi }\Theta _y^2\right] ,
\nonumber \\
\Psi _3 &\simeq &e^{-\imth \theta _x}\exp \left[ -\frac 1{4\pi }\theta
_y^2\right] \exp \left[ -\frac 1{4\pi }\left( \Theta _y-\pi \right)
^2\right] ,
\nonumber \\
\Psi _4 &\simeq &e^{-\imth \theta _x}e^{-\imth \Theta _x}\exp \left[ -\frac 1{4\pi
}\left( \theta _y-\pi \right) ^2 -\frac 1{4\pi }\left(
\Theta _y-\pi \right) ^2\right] .
\nonumber
\end{eqnarray}

Now let's calculate the crystal momentum for the four-fold degenerate ground
states. For the Z2E type mutual \textrm{U(1)}$\times $\textrm{U(1)} CS
theory, the translation operations $T_i$ are known as
\[
T_i^{-1} A_jT_i= a_j,\ \ \ \ \ \ T_i^{-1} a_jT_i= A_j.
\]
Thus we have the translation operation for its zero modes $\left( \Theta
_x,\Theta _y\right) $ and $\left( \theta _x,\theta _y\right) :$
\begin{eqnarray*}
{T}_x^{-1}\theta _x{T}_x &=&\Theta _x, \\
{T}_y^{-1}\theta _y{T}_y &=&\Theta _y, \\
{T}_x^{-1}\theta _y{T}_x &=&\Theta _y, \\
{T}_y^{-1}\theta _x{T}_y &=&\Theta _x.
\end{eqnarray*}
Under the translation operators, we have

\begin{eqnarray*}
{T}_x |  j\rangle &=& |  j\rangle , \\
{T}_y |  j\rangle &=& |  j\rangle , \\
j &=&1,4.
\end{eqnarray*}
\begin{eqnarray*}
{T}_x |  2\rangle &=& |  3\rangle ,{T}_y|  2\rangle
=|  3\rangle , \\
{T}_x |  3\rangle &=& |  2\rangle ,{T}_y|  3\rangle
=|  2\rangle .
\end{eqnarray*}
So $|  2\rangle $ and $|  3\rangle $ cannot be the eigenstates for the
ground state. Instead, the eigenstates for the ground state are given as $
|  2^{\prime }\rangle =\frac 1{\sqrt{2}}\left( |  2\rangle +|
3\rangle \right) $ and $|  3^{\prime }\rangle =\frac 1{\sqrt{2}}\left(
|  2\rangle -|  3\rangle \right) .$ For $|  2^{\prime }\rangle $ and $
|  3^{\prime }\rangle ,$ the eigenvalues for the translation operators are
given as
\begin{eqnarray*}
{T}_x |  2^{\prime }\rangle &=& |  2^{\prime }\rangle ,
{T}_y|  2^{\prime }\rangle =|  2^{\prime }\rangle , \\
{T}_x |  3^{\prime }\rangle &=& e^{\imth \pi }|  3^{\prime }\rangle ,
{T}_y|  3^{\prime }\rangle =e^{\imth \pi }|  3^{\prime }\rangle .
\end{eqnarray*}
As a result, on an even-by-even lattice, the crystal momentum of the E type
mutual \textrm{U(1)}$\times $\textrm{U(1)} CS theory is $(K_x,K_y)=(0,0)$
for the ground states $|  1\rangle ,$ $|  2^{\prime }\rangle $, $|
4\rangle $ and $(K_x,K_y)=(\pi ,\pi )$ for the ground state $|  3^{\prime
}\rangle .$

For other cases, on an even-by-odd, odd-by-even or odd-by-odd lattice, the
situations are changed. Because for odd number rows along \textrm{x}-axis or
\textrm{y}-axis, one gauge field $ A_\mu $ ($ a_\mu $) will turn
into the other one $ a_\mu $ ($ A_\mu $). For example, on a $
L_x\times L_y$ even-by-odd lattice ($L_x$ is an even number and $L_y$ is an
odd number$)$, under such a twisted boundary condition for odd number $L_y$,
one has
\begin{eqnarray}
 A_\mu (x,y+L_y) &=& a_\mu (x,y),
\nonumber \\
 a_\mu (x,y+L_y) &=& A_\mu (x,y),  \nonumber \\
 A_\mu (x+L_x,y) &=& A_\mu (x,y),
\nonumber \\
 a_\mu (x+L_x,y) &=& a_\mu (x,y).  \label{t}
\end{eqnarray}
The quantization for gauge fields in Eq.\ref{modeAa} cannot be applied to
the gauge fields under a twisted boundary condition.

Now after putting the mutual \textrm{U(1)}$\times $\textrm{U(1)} CS
theory on a $L_x\times (2L_y)$ even-by-even lattice, we have a
periodic boundary condition,
\[
 A_\mu (x,y+2L_y)= A_\mu (x,y), a_\mu
(x,y+2L_y)= a_\mu (x,y).
\]
In the temporal gauge, $A_0=0,$ and on such even-by-even lattice, we can
expand the fluctuations for the gauge fields as
\begin{align}
( A_x, A_y)& =(\frac 1{L_x}\Theta _x+\sum_{\mathbf{k}}A_{\mathbf{k}
}^xe^{\imth \check{x}\cdot \mathbf{k}},\frac 1{2L_y}\Theta _y+\sum_{\mathbf{k}}A_{
\mathbf{k}}^ye^{\imth \check{x}\cdot \mathbf{k}}), \\
( a_x, a_y)& =(\frac 1{L_x}\theta _x+\sum_{\mathbf{k}}a_{\mathbf{k}
}^xe^{\imth \check{x}\cdot \mathbf{k}},\frac 1{2L_y}\theta _y+\sum_{\mathbf{k}}a_{
\mathbf{k}}^ye^{\imth \check{x}\cdot \mathbf{k}})
\end{align}
where $\mathbf{k=}(k_x,k_y)=(\frac{2\pi }{L_x}n_x,$ $\frac \pi {L_y}n_y)$
where $n_{x,y}$ are integers. $(A_{\mathbf{k}}^x,A_{\mathbf{k}}^y)$ and $(a_{
\mathbf{k}}^x,a_{\mathbf{k}}^y)$ are the gauge fields with non-zero momentum
and $\left( \Theta _x,\Theta _y\right) $ and $\left( \theta _x,\theta
_y\right) $ are the zero modes with zero momentum for the gauge fields $A_i$
and $a_i$. However, $A_{\mathbf{k}}^i$ and $a_{\mathbf{k}}^i$ ($\Theta _i$
and $\theta _i$ ) are not independent and have constraints to obey the
original twisted boundary condition in Eq.\ref{t}, we must have
\begin{eqnarray}
A_{\mathbf{k}}^i &=&a_{\mathbf{k}}^ie^{iL_y\cdot k_y}=a_{\mathbf{k}
}^ie^{\imth \cdot \pi n_y},  \label{c} \\
\Theta _i &=&\theta _i.  \nonumber
\end{eqnarray}

To calculate the topological degeneracy, we map the original mutual \textrm{
U(1)}$\times $\textrm{U(1)} CS theory on even-by-odd lattice to two-particle
quantum mechanics model on a torus in a magnetic field $\frac 1\pi $. In the
''Landau gauge'', the effective Hamiltonian of the two-particle quantum
mechanics model is given as
\begin{eqnarray*}
H_{eff} &=&\frac{(P_{\Theta _x}-\frac{\theta _y}{2\pi })^2}{2M_x}+\frac{
(p_{\theta _y}+\frac{\Theta _x}{2\pi })^2}{2m_x} \\
&&+\frac{(P_{\Theta _y}+\frac{\theta _x}{2\pi })^2}{2M_y}+\frac{(p_{\theta
_x}-\frac{\Theta _y}{2\pi })^2}{2m_x}
\end{eqnarray*}
where $M_x=\frac 1{e_A^2}\frac{2L_y}{L_x},$ $M_y=\frac 1{e_A^2}\frac{L_x}{
2L_y}$ and $m_x=\frac 1{e_a^2}\frac{2L_y}{L_x},$ $m_y=\frac 1{e_a^2}\frac{L_x
}{2L_y}.$ However, because of the constraint in Eq.\ref{c}, the two
particles $(\theta _x,\theta _y)$ and $(\Theta _x,\Theta _y)$ are bound into
a single particle! As a result, there are two degenerate ground states
instead of four. In addition, we can write down the wave-functions for the
two ground states in the Landau gauge with topological degeneracy: for the
wave-function $|  1\rangle ,$
\[
\Psi _1\simeq e^{-\frac 1{4\pi }\theta _y^2}=e^{-\frac 1{4\pi }\Theta _y^2},
\]
and the wave-function $|  2\rangle $,
\[
\Psi _2\simeq e^{-\imth \Theta _x}e^{-\frac 1{4\pi }\left( \theta _y-\pi \right)
^2}=e^{-\imth \theta _x}e^{-\frac 1{4\pi }\left( \Theta _y-\pi \right) ^2}.
\]

Now let's calculate the crystal momentum for the two-fold degenerate ground
states. The ground states are invariant under the translation operations
\begin{eqnarray*}
{T}_x |  j\rangle =|  j\rangle , \\
{T}_y |  j\rangle =|  j\rangle , \\
j &=&1,2.
\end{eqnarray*}
Then the crystal momentum $(K_x,K_y)$ is $(0,0)$ for the E type mutual
\textrm{U(1)}$\times $\textrm{U(1)} CS theory on an even-by-odd lattice.

Furthermore, using the same method, we calculated the topological
degeneracies and the crystal momenta for the ground states of the Z2E type
mutual \textrm{U(1)}$\times $\textrm{U(1)} CS theory on an odd-by-even or
odd-by-odd lattices. The results are similar to those on an even-by-odd
lattice: the ground states have two-fold degeneracy$\ $and $(K_x,K_y)=(0,0)$.

In summary, all the low energy physical properties for the Z2E type \textrm{
U(1)}$\times $\textrm{U(1)} Chern-Simons theory match that for the Z2E
topological ordered state.

\subsubsection{Topological degeneracy and crystal momenta for the Z2A type
mutual \textrm{U(1)}$\times $\textrm{U(1)} CS theory}

In this part, we will calculate the topological degeneracy and crystal
momenta for Z2A type mutual \textrm{U(1)}$\times $\textrm{U(1)} CS theory.
The effective Hamiltonian to describe the low energy physics of the Z2A type
the mutual \textrm{U(1)}$\times $\textrm{U(1)} CS theory can be written in
the ''Landau gauge'' as
\begin{equation}
H_{eff}=\frac{(P_{\Theta _x}-\frac{\theta _y}\pi )^2}{2M_x}+\frac{p_{\theta
_y}^2}{2m_y}+\frac{(p_{\theta _x}-\frac{\Theta _y}\pi )^2}{2m_x}+\frac{
P_{\Theta _y}^2}{2M_y}.  \nonumber
\end{equation}

It is noted that there exists the Heisenberg Algebra for zero modes of the
gauge fields. The ''magnetic''\ translation operators $U_{\theta _x}=e^{\pi
i({p}_{\theta _x}+\frac{\Theta _y}\pi )}$ and $U_{\Theta _y}=e^{\pi i(
{p}_{\Theta _y}+\frac{\theta _x}\pi )}$ consist of the Heisenberg
algebra
\[
U_{\theta _x}U_{\Theta _y}=e^{\imth \pi }U_{\Theta _y}U_{\theta _x}.
\]
Because the Hamiltonian is invariant for the operations $U_{\theta _x}$ and $
U_{\Theta _y},$
\begin{eqnarray*}
U_{\theta _x}^{-1}HU_{\theta _x} &=&H, \\
U_{\Theta _y}^{-1}HU_{\Theta _y} &=&H,
\end{eqnarray*}
the ground states are the eigenstates of $U_{\theta _x}$ and $U_{\Theta _y}.$
So one can draw a conclusion from the Heisenberg algebra that the ground
states have two-degeneracy for $\left( \theta _x,\Theta _y\right) $. On the
other hand, for $\left( \Theta _x,\theta _y\right) ,$ one can do the same
calculation. So the ground states have two-degeneracy for $\left( \theta
_y,\Theta _x\right) $ which is also characterized by the eigenstates of $
U_{\theta _y}=e^{\pi i({p}_{\theta _y}+\frac{\Theta _x}\pi )}$ and $
U_{\Theta _x}=e^{\pi i({p}_{\Theta _x}+\frac{\theta _y}\pi )}.$ As a
result, for the Z2A type mutual \textrm{U(1)}$\times $\textrm{U(1)} CS
theory, the ground states have four-fold degeneracy :
\begin{equation}
D=D_{(\Theta _x,\theta _y)}D_{(\Theta _y,\theta _x)}=2\times 2=4.
\end{equation}

We denote the four ground states with topological degeneracy as
$|
1\rangle ,$ $|  2\rangle ,$ $|  3\rangle $ and $|  4\rangle ,$
\begin{eqnarray*}
U_{\theta _x} |  1\rangle &=& |  1\rangle , \\
U_{\theta _x} |  2\rangle &=& |  2\rangle , \\
U_{\theta _x} |  3\rangle &=& e^{\imth \pi }|  3\rangle , \\
U_{\theta _x} |  4\rangle &=& e^{\imth \pi }|  4\rangle ,
\end{eqnarray*}
and
\begin{eqnarray*}
U_{\Theta _y} |  1\rangle &=& |  1\rangle , \\
U_{\Theta _y} |  2\rangle &=& e^{\imth \pi }|  2\rangle , \\
U_{\Theta _y} |  3\rangle &=& |  3\rangle , \\
U_{\Theta _y} |  4\rangle &=& e^{\imth \pi }|  4\rangle .
\end{eqnarray*}

Now let's calculate the crystal momentum for the four-fold
degenerate ground states. For the Z2A type mutual
\textrm{U(1)}$\times $\textrm{U(1)} Chern-Simons theory, the
translation operations for the gauge fields are given by Eq.
(\ref{AaTrans}). The translation operations for zero modes of the
gauge fields are given as (\ref{AaTrans})
\begin{eqnarray*}
{T}_x^{-1}\Theta _y{T}_x &=&\Theta _y, \\
{T}_y^{-1}\Theta _x{T}_y &=&\Theta _x, \\
{T}_x^{-1}\Theta _y{T}_x &=&\Theta _y+L_y\pi , \\
{T}_y^{-1}\Theta _x{T}_y &=&\Theta _x,
\end{eqnarray*}
and
\begin{eqnarray*}
{T}_x^{-1}\theta _y{T}_x &=&\theta _y, \\
{T}_y^{-1}\theta _x{T}_y &=&\theta _x+L_x\pi , \\
{T}_x^{-1}\theta _x{T}_x &=&\theta _x, \\
{T}_y^{-1}\theta _y{T}_y &=&\theta _y.
\end{eqnarray*}
As a result, the real ground states can be labeled by the eigenvalues of $
U_{\theta _x}$ (or $U_{\theta _y},$ $U_{\Theta _y},$ $U_{\Theta _x}$) which
are $1$ and $-1$. We denote the four ground states with topological
degeneracy as $|  1\rangle ,$ $|  2\rangle ,$ $|  3\rangle $ and $|
4\rangle ,$
\begin{eqnarray*}
U_{\theta _x} |  1\rangle &=& |  1\rangle , \\
U_{\theta _x} |  2\rangle &=& |  2\rangle , \\
U_{\theta _x} |  3\rangle &=& e^{\imth \pi }|  3\rangle , \\
U_{\theta _x} |  4\rangle &=& e^{\imth \pi }|  4\rangle .
\end{eqnarray*}

Firstly, on an even-by-even lattice, the translation operations for its zero
modes lead to trivial results
\begin{eqnarray*}
{T}_x^{-1}\Theta _y{T}_x &=&\Theta _y, \\
{T}_y^{-1}\Theta _x{T}_y &=&\Theta _x, \\
{T}_x^{-1}\Theta _y{T}_x &=&\Theta _y, \\
{T}_y^{-1}\Theta _x{T}_y &=&\Theta _x.
\end{eqnarray*}
and
\begin{eqnarray*}
{T}_x^{-1}\theta _y{T}_x &=&\theta _y, \\
{T}_y^{-1}\theta _x{T}_y &=&\theta _x, \\
{T}_x^{-1}\theta _x{T}_x &=&\theta _x, \\
{T}_y^{-1}\theta _y{T}_y &=&\theta _y,
\end{eqnarray*}
From them, we have
\begin{eqnarray*}
{T}_x |  j\rangle &=& |  j\rangle , \\
{T}_y |  j\rangle &=& |  j\rangle , \\
j &=&1,2,3,4.
\end{eqnarray*}
Then the crystal momentum $(K_x,K_y)$ of the four-fold degenerate ground
states $|  j\rangle $ is $(0,0).$

Secondly on an odd by even lattice ($L_x$ is odd number and $L_y$ is even
number), the translation operations are given as
\begin{eqnarray*}
{T}_x^{-1}\theta _y{T}_x &=&\theta _y, \\
{T}_y^{-1}\theta _x{T}_y &=&\theta _x+\pi , \\
{T}_x^{-1}\theta _x{T}_x &=&\theta _x, \\
{T}_y^{-1}\theta _y{T}_y &=&\theta _y,
\end{eqnarray*}
and
\begin{eqnarray*}
{T}_x^{-1}\Theta _y{T}_x &=&\Theta _y, \\
{T}_y^{-1}\Theta _x{T}_y &=&\Theta _x, \\
{T}_x^{-1}\Theta _x{T}_x &=&\Theta _x, \\
{T}_y^{-1}\Theta _y{T}_y &=&\Theta _y.
\end{eqnarray*}
Now the translation operator ${T}_y$ turns into the ''magnetic''\
translation operator $U_{\theta _x}=e^{\pi i({p}_{\theta _x}+\frac{
\Theta _y}\pi )},$
\[
{T}_y|  \imth \rangle =U_{\theta _x}|  \imth \rangle =e^{\pi i({p}_{\theta
_x}+\frac{\Theta _y}\pi )}|  \imth \rangle \text{, }i=1,2,3,4.
\]

Under the translation operations on the wave functions in Eq.\ref{wave}, we
have
\begin{eqnarray*}
{T}_x |  1\rangle &=& |  1\rangle , \\
{T}_x |  2\rangle &=& |  2\rangle , \\
{T}_x |  3\rangle &=& |  3\rangle , \\
{T}_x |  4\rangle &=& |  4\rangle ,
\end{eqnarray*}
and
\begin{eqnarray*}
{T}_y |  1\rangle &=& U_{\theta _x}|  1\rangle =|  1\rangle , \\
{T}_y |  2\rangle &=& U_{\theta _x}|  2\rangle =|  2\rangle
, \\
{T}_y |  3\rangle &=& U_{\theta _x}|  2\rangle =e^{\imth \pi }|
3\rangle , \\
{T}_y |  4\rangle &=& U_{\theta _x}|  2\rangle =e^{\imth \pi }|
4\rangle .
\end{eqnarray*}
Using the same method, we can obtain that the crystal momentum of the two
ground states $|  1\rangle $ and $|  2\rangle $ is $(0,0).$ The crystal
momentum of the other two ground states $|  3\rangle $ and $|  4\rangle $
is $(0,\pi ).$

Thirdly, on an even-by-odd lattice ($L_x$ is even number and $L_y$ is odd
number), the translation operations for its zero modes lead to non-trivial
results
\begin{eqnarray*}
{T}_x^{-1}\Theta _y{T}_x &=&\Theta _y, \\
{T}_y^{-1}\Theta _x{T}_y &=&\Theta _x, \\
{T}_x^{-1}\Theta _y{T}_x &=&\Theta _y+\pi , \\
{T}_y^{-1}\Theta _x{T}_y &=&\Theta _x,
\end{eqnarray*}
and
\begin{eqnarray*}
{T}_x^{-1}\theta _y{T}_x &=&\theta _y, \\
{T}_y^{-1}\theta _x{T}_y &=&\theta _x, \\
{T}_x^{-1}\theta _x{T}_x &=&\theta _x, \\
{T}_y^{-1}\theta _y{T}_y &=&\theta _y.
\end{eqnarray*}
Then the translation operator ${T}_x$ turns into the ''magnetic''\
translation operator $U_{\Theta _y}=e^{\pi i({p}_{\Theta _y}+\frac{
\theta _x}\pi )},$
\[
{T}_x|  \imth \rangle =U_{\Theta _y}|  \imth \rangle =e^{\pi i({p}_{\Theta
_y}+\frac{\theta _x}\pi )}|  \imth \rangle \text{, }i=1,2,3,4.
\]
From them, we have
\begin{eqnarray*}
{T}_x |  1\rangle &=& U_{\Theta _y}|  1\rangle =|  1\rangle ,\text{
} \\
{T}_x |  2\rangle &=& U_{\Theta _y}|  2\rangle =e^{\imth \pi }|
2\rangle , \\
{T}_x |  3\rangle &=& U_{\Theta _y}|  3\rangle =|  3\rangle ,\text{
} \\
{T}_x |  4\rangle &=& U_{\Theta _y}|  4\rangle =e^{\imth \pi }|
4\rangle ,
\end{eqnarray*}
and
\begin{eqnarray*}
{T}_y |  1\rangle &=& |  1\rangle , \\
{T}_y |  3\rangle &=& |  3\rangle , \\
{T}_y |  2\rangle &=& |  2\rangle , \\
{T}_y |  4\rangle &=& |  4\rangle .
\end{eqnarray*}
The crystal momentum of two ground states $|  1\rangle $ and $|
3\rangle $ is $(0,0).$ The crystal momentum of the other two ground states $
|  4\rangle $ and $|  2\rangle $ is $(\pi ,0).$

Fourthly for $L_x$ and $L_y$ are all odd numbers (on an odd-by-odd lattice),
the translation operations become
\begin{eqnarray*}
{T}_x^{-1}\Theta _y{T}_x &=&\Theta _y+\pi , \\
{T}_y^{-1}\Theta _x{T}_y &=&\Theta _x, \\
{T}_x^{-1}\Theta _x{T}_x &=&\Theta _x, \\
{T}_y^{-1}\Theta _y{T}_y &=&\Theta _y,
\end{eqnarray*}
and
\begin{eqnarray*}
{T}_x^{-1}\theta _y{T}_x &=&\theta _y, \\
{T}_y^{-1}\theta _x{T}_y &=&\theta _x+\pi , \\
{T}_x^{-1}\theta _x{T}_x &=&\theta _x, \\
{T}_y^{-1}\theta _y{T}_y &=&\theta _y.
\end{eqnarray*}
Then the translation operators ${T}_x$ and ${T}_y$ turn into the
''magnetic''\ translation operator $U_{\Theta _y}=e^{\pi
i({p}_{\Theta _y}+\frac{\theta _x}\pi )}$ and $U_{\theta _x}=e^{\pi
i({p}_{\theta _x}+ \frac{\Theta _y}\pi )},$
\begin{eqnarray*}
{T}_x |  \imth \rangle &=& U_{\Theta _y}|  \imth \rangle =e^{\pi i({p}
_{\Theta _y}+\frac{\theta _x}\pi )}|  \imth \rangle \text{, } \\
{T}_y |  \imth \rangle &=& U_{\theta _x}|  \imth \rangle =e^{\pi i({p}
_{\theta _x}+\frac{\Theta _y}\pi )}|  \imth \rangle \text{, }i=1,2,3,4.
\end{eqnarray*}
Now ${T}_x$ and ${T}_y$ must obey the Heisenberg algebra for $
U_{\Theta _y}$ and $U_{\theta _x}$
\begin{equation}
{T}_x{T}_y=e^{\imth \pi }{T}_y{T}_x.  \label{an}
\end{equation}
On the other hand, the translation symmetry of the system leads to the
commutation relationship between ${T}_x$ and ${T}_y$
\begin{equation}
{T}_x{T}_y={T}_y{T}_x.  \label{com}
\end{equation}
The only solve to the Eq.\ref{an} and Eq.\ref{com} is $|  \imth \rangle \equiv
0 $. That is, there don't exist the four degenerate ground states at all. We
can see that for the real ground states, the $A_\mu $ and $a_\mu $ charges
for the excitations cannot be zero on an odd by odd lattice. So the non zero
background charge leads to an infinity degeneracy on odd by odd lattice for
the Z2A type mutual \textrm{U(1)}$\times $\textrm{U(1)} CS theory.

As a result, all the low energy physical properties for the Z2A type \textrm{
U(1)}$\times $\textrm{U(1)} Chern-Simons theory match that for the Z2A
topological ordered state.

\bibliographystyle{apsrev}

\end{document}